\documentclass[aps,pre,preprint,showpacs]{revtex4}
\usepackage{setspace}
\usepackage{amsmath}
\usepackage{amssymb}
\usepackage{epsfig}
 \normalsize
\begin{document}
\preprint{}
\title{The
isotropic-cholesteric phase transition of filamentous virus
suspensions as a function of rod length and charge}
\author{Kirstin Purdy}
\affiliation{Complex Fluids Group, Department of Physics, Brandeis
University, Waltham, Massachusetts 02454}
\author{Seth Fraden}
\affiliation{Complex Fluids Group, Department of Physics, Brandeis
University, Waltham, Massachusetts 02454}
\date{\today}
\begin{abstract}
The viruses studied are genetically engineered, charged,
semiflexible filamentous bacteriophages that are structurally
identical to M13 virus, but differ either in contour length or
surface charge. While varying contour length ($L$) we assume the
persistence length ($P$) remains constant, and thus we alter the
rod flexibility ($L/P$). Surface charge is altered both by
changing solution pH and by comparing two viruses, {\it fd} and
M13, which differ only by the substitution of one charged for one
neutral amino acid per virus coat protein. We measure both the
isotropic and cholesteric coexistence concentrations as well as
the nematic order parameter after unwinding the cholesteric phase
in a magnetic field. The isotropic-cholesteric transition
experimental results agree semi-quantitatively with theoretical
predictions for semiflexible, charged rods.
\end{abstract}

 \maketitle

\section{Introduction}
For a suspension of rigid rodlike particles, Onsager determined
that hard-core interactions alone are sufficient for inducing an
entropy driven phase transition from an isotropic phase, in which
the particles are randomly oriented, to a nematic phase, in which
the orientation of the particles is distributed about a preferred
direction\cite{Onsager49}. When the rodlike particles are
semiflexible and/or charged, like many biopolymers such as DNA and
F-Actin, the properties of the phase transition can differ
significantly from those predicted for hard, rigid rods. Small
amounts of flexibility are predicted \cite{Khokhlov82} and
observed \cite{Tang95,Sato96} to increase the stability of the
isotropic phase and lead to a less ordered nematic phase. In this
paper, we study the effects of flexibility on the
isotropic-nematic (I-N) transition using suspensions of the
rodlike charged, semiflexible M13 virus and M13 virus
length-mutants. By varying the contour length ($L$) of our
experimental charged rods while maintaining a constant persistence
length ($P$) we change the rod flexibility ($L/P$). The
persistence length is defined as the length over which tangent
vectors along a polymer are correlated \cite{Grosberg97}. In our
experiments the flexibility of the rods remains within the
semiflexible limit, where $P\sim L$. The effect of surface charge
on the I-N transition of charged rods is also investigated.
Surface charge is varied by modifying both the surface chemistry
of the rods and the solution chemistry, by changing pH.

While Onsager developed the original theory for the
isotropic-nematic transition of hard and charged rigid rodlike
particles, Khokhlov and Semenov were responsible for incorporating
flexibility into this theory \cite{Khokhlov82}. They extended
Onsager's theory to include systems of semiflexible rods with a
large length ($L$) to diameter ($D$) aspect ratio ($L/D$) and
arbitrary persistence length. They explicitly calculated the
equilibrium properties of the I-N phase transition in the limit of
very flexible $L/P>>1$ and very rigid $L/P<<1$ rods and
interpolated between the two limits to find the properties of
semiflexible rod phase behavior. Shortly afterwards, Chen
numerically calculated the concentrations of the coexisting
isotropic and nematic phases as well as the order parameter of the
coexisting nematic phase for arbitrary flexibility using
Khokhlov-Semenov theory\cite{Chen93}. For rigid rods, the limit of
stability of the isotropic phase is predicted to be $c_i=4/b$,
where $c_i$ is the number density and $b=\pi L^2 D/4$, the average
excluded volume in the isotropic phase \cite{Kayser78}. For
flexible rods, Khokhlov-Semenov theory predicts that slight
semiflexibility will increase the stability of the isotropic phase
by increasing $bc_i$, and will narrow the I-N coexistence region.
Flexibility is also predicted to significantly lower the nematic
order parameter at coexistence. The nematic order parameter $S$ is
the second moment of the orientational distribution function of
the rods, $f(\theta)$, or $S=2\pi\int
P_2(\cos(\theta))f(\theta)d\theta$, where $P_2$ is the second
Legendre polynomial. For a completely aligned nematic $S=1$,
whereas for an isotropic phase $S=0$. For rigid rods the predicted
nematic order parameter at coexistence is $S= 0.79$
\cite{Herzfeld84}. The predictions from the Khokhlov-Semenov
theory show quantitative agreement with the measured I-N
transition for suspensions of charged semiflexible virus {\it fd},
charged polymer xanthan, and neutral polymer PBLG
\cite{Tang95,Sato96}.

\begin{figure}
\centerline{\epsfig{file=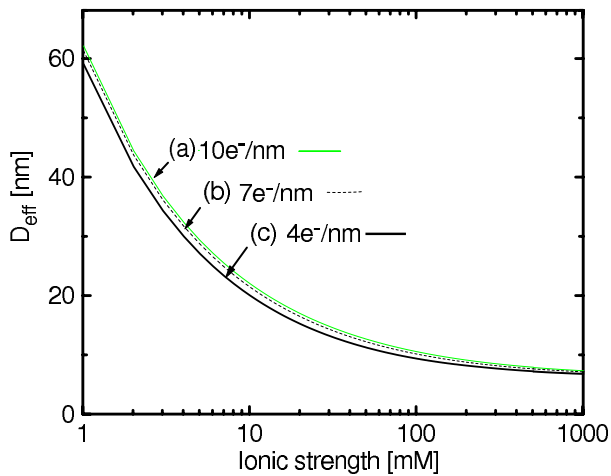,
width=8cm}}\caption[Effective diameter calculation as a function
of ionic strength and surface charge]{\label{Deff.fig}(Color
Online) Effective diameter as a function of ionic strength and
surface charge. With increasing ionic strength
$D_{\mbox{\scriptsize eff}}$ approaches the bare diameter of {\it
fd} (M13) $D=6.6$ nm. The effective diameter is plotted for
surface charges of 10 e$^-$/nm, 7 e$^-$/nm and 4 e$^-$/nm. These
surface charge densities are the same as those measured for (a)
{\it fd} at pH 8.2, (b) M13 at pH 8.2 or {\it fd} at pH 5.2, and
(c) M13 at pH 5.2. At these surface charge densities
$D_{\mbox{\scriptsize eff}}$ is insensitive to variation in
charge.}
\end{figure}

Electrostatic interactions are incorporated into the Onsager model
by rescaling the bare rod diameter $D$ to a larger effective
diameter $D_{\mbox{\scriptsize{eff}}}$ which depends on the ionic
properties of the particle and the solution \cite{Onsager49,
Stroobants86a}. $D_{\mbox{\scriptsize{eff}}}$ is calculated from
the second virial coefficient of Onsager's free energy equation
for charged rigid rods. In Fig. \ref{Deff.fig} we plot
$D_{\mbox{\scriptsize eff}}$ as described by Stroobants {\it et
al.} \cite{Stroobants86a} as functions of ionic strength and rod
surface charge. The non-linear Poisson-Boltzmann equation used in
Stroobants description of $D_{\mbox{\scriptsize eff}}$ was solved
numerically using the approximations developed by Philip and
Wooding \cite{Philip70}. With increasing ionic strength
$D_{\mbox{\scriptsize eff}}$ decreases approaching the bare rod
diameter. Past experiments have shown that $D_{\mbox{\scriptsize
eff}}$ accurately describes the ionic strength dependence of the
I-N transition of {\it fd} virus suspensions \cite{Tang95}. For
highly charged rods, the effect of surface charge on
$D_{\mbox{\scriptsize{eff}}}$ is small as the non-linear nature of
the Poisson-Boltzmann equation leads to counterion condensation
near the colloid surface which renormalizes the bare surface
charge to a lesser effective charge, which is nearly independent
of the bare surface charge. In the nematic phase, the effective
diameter increases due to an added effect called ``twist" which is
characterized by the parameter $h=\kappa
^{-1}/D_{\mbox{\scriptsize{eff}}}$, where $\kappa^{-1}$ is the
Debye screening length. The effect of twist on
$D_{\mbox{\scriptsize{eff}}}$, however, is predicted to be small
for {\it fd} \cite{Tang95}, and we neglect it here. Studying the
influence of ionic strength and surface charge on the I-N phase
behavior tests if $D_{\mbox{\scriptsize{eff}}}$ can be accurately
used to map charged rod phase behavior to hard-rod theories.

Onsager's theory is based on an expansion of the free energy
truncated at the second virial level, so that only two-particle
interactions are considered. This assumption has been shown to be
accurate in the limit of very long rods, where $L/D>100$
\cite{Straley73}, or for very dilute suspensions. In our
experimental system, however, decreasing the ionic strength
rapidly decreases our effective aspect ratio to values far below
the $L/D=100$ limit. In order to accurately predict the phase
behavior of rods with an effectively small aspect ratio, the
theoretical free energy needs to incorporate third and higher
virial coefficients. Scaled particle theory (SPT), which
incorporates all higher virial coefficients in an approximate way
is one theory which accomplishes this\cite{Cotter79}. A scaled
particle theory for hard rigid rods was originally developed by
Cotter \cite{Cotter79}. More recently we have expanded this theory
to include charge and semiflexibility \cite{Purdy03,Dogic04}. In
conjunction with the Khokhlov-Semenov second virial theory we use
this scaled particle theory to interpret our experimental results.

In this paper we present experimental measurements of the
isotropic-nematic phase transition of semiflexible charged
colloidal rods as a function of rod length, surface composition,
solution pH, and solution ionic strength. We measure both the
coexistence concentrations and the nematic order parameter and
compare our results to both Onsager's theory, by way of Chen's
numerical calculation\cite{Chen93}, and scaled particle theory.
For our model rods we use monodisperse suspensions of charged
semiflexible rodlike {\it fd} virus, wild type M13 virus, and
mutants of M13 virus which differ from the wild type only by their
contour length. In solution, these particles exhibit isotropic,
cholesteric (or chiral nematic), and smectic phases
\cite{Lapointe73,Tang95,Dogic97,Dogic01}. Suspensions of {\it fd}
have been previously shown to exhibit an I-N transition which
agrees with theoretical predictions for semiflexible rods with an
effective diameter $D_{\mbox{\scriptsize eff}}$ \cite{Tang95}. M13
virus is structurally identical to {\it fd} virus, differing only
in surface charge, making these two particles an ideal system for
studying the influence of bare surface charge on the
isotropic-nematic transition. Additionally, by comparing the I-N
phase behavior of each of the M13 mutants, which except for length
are structurally identical, and therefore by assumption have the
same persistence length, we measure the influence of flexibility,
defined as the ratio $L/P$ on this transition. Though {\it fd} and
M13 exhibit a cholesteric phase, the free energy difference
between the cholesteric and the nematic phase is much smaller than
the difference between the isotropic and nematic phases
~\cite{deGennes93}. This allows us to compare our results to
theoretical predictions for the isotropic-nematic (I-N)
transition. We refer to the cholesteric phase as the nematic phase
henceforth.

Motivation for these length and surface charge dependent
measurements of the I-N transition arose because new measurements
of the nematic-smectic (N-S) transition in this same system
\cite{Purdy04sm} exhibit measurable surface charge dependence and
ionic strength dependence which can not be accounted for by
treating the virus as a hard rod with a diameter
$D_{\mbox{\scriptsize eff}}$, in contrast to our previous
measurements, which were limited in range of ionic strength
\cite{Dogic97}. The new N-S measurements inspired a closer look at
the ability of $D_{\mbox{\scriptsize eff}}$ to describe the
effects of surface charge on the I-N transition. New measurements
of the N-S transition as a function of length also indicate that
semiflexibility has no measurable effect on the N-S transition for
the limited range studied, which is as predicted, but which is in
sharp contrast to the large predicted effect of flexibility on the
I-N transition for the same range. The measurements presented here
of the I-N transition as a function of charge and flexibility will
contribute to the understanding of the relative importance of
these variables in the evolution of the liquid crystalline
ordering of charged semiflexible rodlike particles with
concentration.

\section{Materials and Methods}
Properties of {\it fd} and wild type M13 include length $L=$0.88
$\mu$m, diameter $D=6.6$ nm, persistence length $P=$2.2 $\mu$m and
molecular weight $M=1.64\times10^7$ g/mol\cite{Fraden95}. Each
virus consists of approximately 2700 coat proteins helicoidally
wrapped around single stranded DNA. The two viruses differ only by
one amino acid per coat protein. In {\it fd} this amino acid is
the negatively charged aspartate (asp$_{12}$), and in M13 it is
the neutral asparagine (asn$_{12}$)\cite{Marvin94}. Thus at near
neutral pH {\it fd} has one more negative charge per coat protein
($3.4\pm 0.1$ e$^-$/protein) than M13 ($2.3\pm 0.1$
e$^-$/protein), which results in a net charge difference of
approximately 30\% \cite{Zimmermann86}. X-ray diffraction studies
are unable to clearly discern any structural differences between
M13 and {\it fd} \cite{Glucksman92}. The M13 length-mutants share
the same properties as wild type M13, varying only in length and
molecular weight, which scales linearly with length. The M13
mutant have lengths of 1.2$\mu$m, 0.64$\mu$m, and 0.39$\mu$m. Wild
type M13, {\it fd}, and M13K07 (the 1.2$\mu$m mutant phage) were
grown using standard techniques \cite{Maniatis89}. The other two
mutant phages were grown using the phagemid method, which produces
bidisperse solutions of the phagemid and the M13K07 helper phage
\cite{Maniatis89}. We chose two plasmid DNA sequences, PGTN28
(4665bp) and LITMUS38 (2820bp) (New England Biolabs, Cambridge MA)
to form our phagemids of length 0.64$\mu$m and 0.39$\mu$m,
respectively. Sample polydispersity was checked using gel
electrophoresis on the intact virus, and on the viral DNA. Except
for the phagemid solutions, which contained approximately 20\% by
mass helper phage M13K07, the virus solutions were highly
monodisperse as indicated by sharp electrophoresis bands.

In a bidisperse system of long and short rods it is predicted that
when isotropic and nematic phases are in coexistence, the longer
rods will strongly partition into the nematic phase
\cite{Lekkerkerker84,Sato94}. Using this fractionation effect we
attempted to purify the bidisperse suspensions of the phagemid and
M13K07 helper phage. We observed partitioning of the long rods
into the nematic phase by DNA agarose gel electrophoresis (2-3
fold more long rods in the cholesteric phase than in the isotropic
phase in qualitative agreement with Lekkerkerker {\it et al.}
\cite{Lekkerkerker84}), but were unable to successfully measure a
difference in long rod concentrations in the isotropic phase after
successive iterations of fractionation. The effect of
fractionation on the coexistence concentrations was assayed by
comparing the isotropic and nematic concentrations of coexisting
samples (about 50\% of each phase in one sample) with the highest
concentrations for which the samples remained completely isotropic
and the lowest concentrations for which the samples remained
completely nematic, respectively. The only difference we observed
was that the nematic concentration measured in coexistence with
the isotropic phase was consistently about 5-10\% lower than the
nematic concentration measured when the sample was 100\% nematic.
The lower concentrations in the coexisting nematic phases are due
to the partitioned long rods undergoing the I-N phase transition
at lower mass concentrations. Because the effect of bidispersity
is small, we report the phase behavior for the 0.39$\mu$m and 0.64
$\mu$m rods at the limits of the coexistence region with the
understanding that the samples contain about $\sim$20\% (by mass)
1.2 $\mu$m rods.

\begin{table}
\begin{tabular}{|l|c|c|c|c|c|c|} \hline
&A&B&C&D&E&F\\
\hline pH&{\it fd}&M13 e$^-$/subunit& mobility ratio& M13 e$^-$/subunit&{\it fd}& M13 \\
&e$^-$/subunit&(charge of {\it fd} minus 1)&$m_{\mbox{\scriptsize M13}}$/$m_{fd}$&(electrophoresis)&e$^-$/nm&e$^-$/nm\\
\hline 8.2&3.4$\pm ~0.1$ &2.4$\pm~ 0.1$&0.67&2.3$\pm~ 0.05$&10&7\\
\hline 5.2&2.3$\pm ~0.1$&1.3$\pm~0.1$&0.5& 1.2$\pm ~0.05$&7&3.6\\
\hline
\end{tabular}
\caption[Table of virus surface charge at two values of solution
pH]{\label{table1} Surface charge of {\it fd} and M13 at pH 8.2
and 5.2. (A) The charge of {\it fd} obtained by titration
experiments \cite{Zimmermann86}. (B) M13 has one less negative
amino acid per coat protein than {\it fd}, thus the surface charge
of M13 can be approximated by subtracting one charge per protein
subunit from the {\it fd} surface charge values. (C) Ratio of
electrophoretic mobility ($m$), determined from Fig. \ref{gel}, of
M13 to {\it fd}. (D) By multiplying the known {\it fd} charge by
$m$, the linear surface charge density of M13 can be calculated.
(E),(F) {\it fd} and M13 surface charge per unit length,
respectively. }
\end{table}

\begin{figure}\centerline{\epsfig{file=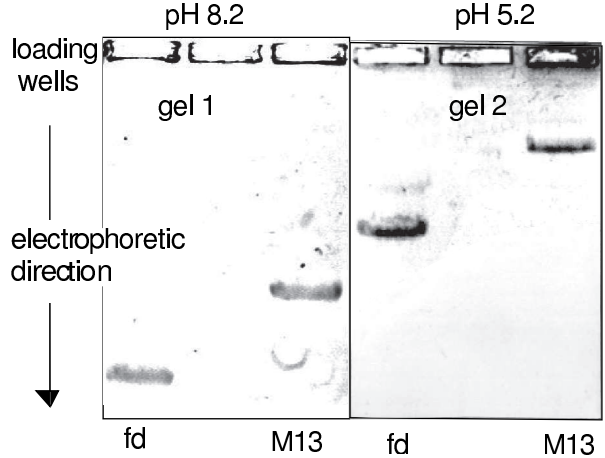,width=8cm}}
\caption[Image of whole M13 and {\it fd} virus gel
electrophoresis.] {\label{gel}Agarose gel electrophoresis of {\it
fd} and M13 virus at pH 8.2 (gel 1) and pH 5.2 (gel 2). At pH 5.2
the buffer was 40 mM Sodium Acetate, and at pH 8.2 the buffer was
40 mM Tris-Acetate-EDTA (TAE). Gels were run at $\sim$1.0\%
agarose concentration and $\sim$3.5 V/cm for 4 hours. Samples were
placed in loading wells at a concentration of approximately 0.3
mg/ml. M13 and {\it fd} have the same length ($L=0.88\mu$m) and
diameter ($D=6.6$nm), and differ only in surface charge. The ratio
of electrophoretic migration distances between M13 and {\it fd}
within each gel is therefore equal to the ratio of the surface
charge. The electrophoresis bands for fd at pH 5.2 and M13 at pH
8.2 are not at the same migration distance, because the absolute
migration distance is also a function of the buffer ions.}
\end{figure}

All samples were dialyzed against a 20mM Tris-HCl buffer at pH 8.2
or 20mM Sodium Acetate buffer adjusted with Acetic Acid to pH 5.2.
To vary ionic strength, NaCl was added to the buffering solution.
The values for surface charge of {\it fd} and M13 at pH 8.2 and pH
5.2 are presented in Table \ref{table1}. The surface charge of
{\it fd} was determined by titration experiments
\cite{Zimmermann86}, and the surface charge of M13 was calculated
in two ways, both starting from the known {\it fd} surface charge.
One way is to compare the molecular composition of {\it fd} and
M13, and the second is to use the fact that because M13 and {\it
fd} are identical except for their surface charge, their the
electrophoretic mobilities are proportional to net surface charge
~\cite{Overbeek50}. In Fig. \ref{gel}, we show using agarose gel
electrophoresis of intact virus that {\it fd} migrates 200\%
faster than M13 at pH 5.2 and 150\% faster at pH 8.2. Note in
Table \ref{table1} we show that the surface charge of M13 at pH
8.2 is the same as the surface charge of {\it fd} at pH 5.2.

All measurements were done at room temperature. The virus
concentrations were measured by absorption spectrophotometry with
the optical density ($A$) of the virus being
$A_{\mbox{\scriptsize{269nm}}}^{\mbox{\scriptsize{1mg/ml}}}=3.84$
for a path length of 1 cm. The nematic order parameter was
obtained by unwinding and aligning the cholesteric phase in a 2T
permanent magnet (SAM-2 Hummingbird Instruments, Arlington, MA
02474)\cite{Oldenbourg86} and measuring the sample birefringence.
At 2T, the magnetic field has a negligible effect on nematic
ordering\cite{Torbet81,Tang93}. The nematic order parameters were
calculated from the optical birefringence measurements obtained
with a Berek compensator using the equation $\Delta
n_{\mbox{\scriptsize sat}}S = \Delta n$, where $\Delta
n_{\mbox{\scriptsize sat}}$ is the saturation birefringence. The
value for $\Delta n_{\mbox{\scriptsize sat}}/\rho =3.8\times
10^{-5}$[ml/mg], where $\rho$ is the concentration of virus in
[mg/ml], as determined for {\it fd} via x-ray diffraction
\cite{Purdy03}.

\section{Results}
\subsection{Effect of length and flexibility on the isotropic-nematic transition}
\begin{figure}
\centerline{\epsfig{file=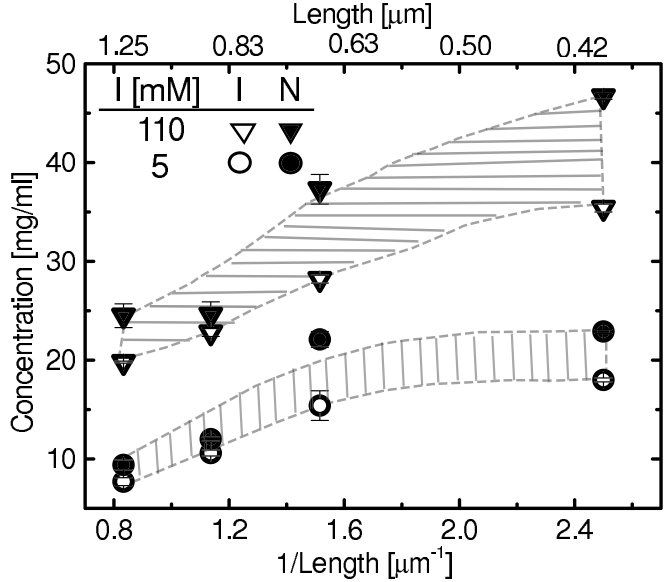,width=8cm}}\caption[Isotropic-nematic
phase transition as a function of particle length and ionic
strength. ]{\label{INmutant.fig} Isotropic-nematic coexistence
concentrations as a function of M13 mutant contour length at 5 mM
and 110 mM ionic strengths at pH 8.2. Open symbols represent the
coexisting isotropic phase and solid symbols the nematic phase.
Shaded areas are a guide to the eye indicating the coexistence
regions. For rigid rods the coexistence concentrations $\rho_i
\propto 1/L $ at a constant ionic strength (constant
$D_{\mbox{\scriptsize eff}}$). Deviations from this relationship
are most likely due to rod flexibility.}
\end{figure}

\begin{figure}
\centerline{\epsfig{file=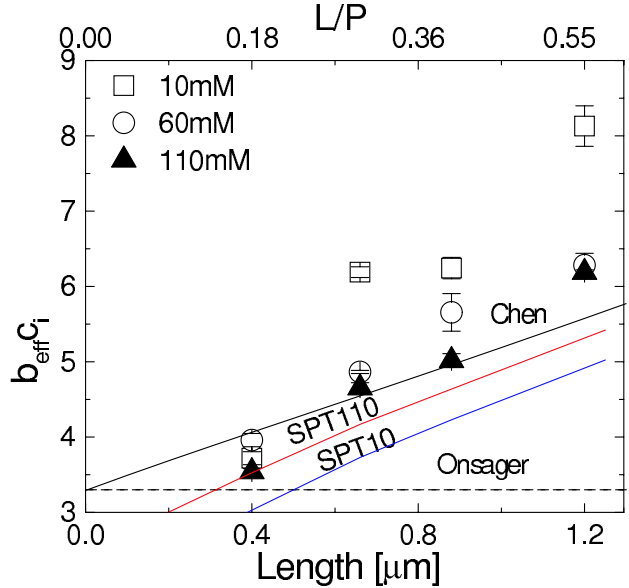,width=8cm}}\caption[Dimensionless
isotropic coexistence concentration as a function of particle
length. ]{\label{bc.fig} Dimensionless concentration of the
isotropic phase in coexistence with the nematic phase as a
function of M13 mutant contour length for three ionic strengths at
pH 8.2. The concentration is defined as $b_{\mbox{\scriptsize
eff}} c_i=\frac{\pi}{4}D_{\mbox{\scriptsize eff}}L^2N_i/V=
25\rho_i\mbox{[mg/ml]}L[\mu\mbox{m}]D_{\mbox{\scriptsize
eff}}$[$\mu$m]. Scale on the top of the graph identifies the
flexibility in terms of $L/P$ with $P=2.2\mu$m. If the rods are
rigid the phase behavior is predicted to be independent of length
(Onsager) (dashed line). Semi-flexible rods show increasing
$b_{\mbox{\scriptsize eff}}c_i$ with increasing flexibility as
predicted by Khokhlov-Semenov theory calculated by Chen (solid
line). Scaled particle theory at 100 mM ionic strength (SPT110)
and at 10 mM ionic strength (SPT10) indicate that
$b_{\mbox{\scriptsize eff}}c_i$ depends on $L/D_{\mbox{\scriptsize
eff}}$.}
\end{figure}

Figure \ref{INmutant.fig} presents the length dependence of the
I-N coexistence concentrations at high (110 mM) and low (5 mM)
ionic strength. For rigid rods $b_{\mbox{\scriptsize eff}}c_i$,
the dimensionless concentration of the isotropic phase in
coexistence with the nematic phase, is predicted to be a constant,
$b_{\mbox{\scriptsize eff}}c_i=3.29$ \cite{Kramer98}, where
$b_{\mbox{\scriptsize eff}}=\frac{\pi}{4}L^2D_{\mbox{\scriptsize
eff}}$ and $c_i=\rho_i N_A/ M$. In $c_i$, $\rho_i$ is the
isotropic mass density, $N_A$ is Avogadro's number, and $M$ is the
molecular weight. Because the molecular weight is proportional to
viral length, $M=M_{wt} L/L_{wt}$, with $M_{wt}$ and $L_{wt}$
equal to the molecular weight and length of wild type M13. Thus $
b_{\mbox{\scriptsize eff}}c_i=  \rho_i L D_{\mbox{\scriptsize
eff}}(\frac{\pi}{4}L_{wt}N_A/M )=
25\rho_i\mbox{[mg/ml]}L[\mu\mbox{m}]D_{\mbox{\scriptsize
eff}}$[$\mu$ m]. Therefore, for rigid rods, $\rho_i
=\mbox{const}/L/D_{\mbox{\scriptsize eff}}$, and at constant ionic
strength (constant $D_{\mbox{\scriptsize eff}}$) $\rho_i$ should
be proportional to $1/L$. However, we observe that at a given
ionic strength, the slope of $\rho_i$ vs $1/L$ is not linear in
Fig. \ref{INmutant.fig}, but instead increases with rod length,
corresponding to an increase in $b_{\mbox{\scriptsize eff}}c_i$.
This is shown more clearly in Fig. \ref{bc.fig}, where
$b_{\mbox{\scriptsize eff}}c_i$ is plotted as a function of
length. The increase in $b_{\mbox{\scriptsize eff}}c_i$ with
length is in agreement with predictions for rods of increasing
flexibility ($L/P$), as shown by the theoretical curves from
Khokhlov-Semenov theory and from SPT for semiflexible rods with a
persistence length of $P=2.2\mu$m. At high ionic strength ( I $>$
60 mM) we see good agreement with Khokhlov-Semenov theory
calculated numerically by Chen (solid line)~\cite{Chen93}. However
with decreasing ionic strength, we measure an increase in the
flexibility dependence of $b_{\mbox{\scriptsize eff}}c_i$.
Subsequently, Khokhlov-Semenov theory only qualitatively describes
the experimental results at low ionic strength. Agreement of the
hard-rod Khokhlov-Semenov theory with our data is better at high
ionic strength than at low ionic strength because the range of
electrostatic interactions is weaker and $L/D_{\mbox{\scriptsize
eff}}$ is large, making the second virial approximation valid.

To interpret the observed increase in flexibility dependence of
the phase transition with decreasing ionic strength, we turn to
the scaled particle theory. The method for determining the scaled
particle theoretical coexistence concentrations and nematic order parameters is described 
elsewhere \cite{Dogic04}. In Fig. \ref{bc.fig} we present the
predicted SPT isotropic coexistence concentrations for rods with a
diameter of $10.4$ nm, (110 mM ionic strength), and $29.4$ nm (10
mM ionic strength). At high ionic strength, SPT shows fair
agreement with experimental results, and the theoretical curve for
$b_{\mbox{\scriptsize eff}}c_i$ is close to that predicted by Chen
for the infinitely long rod limit. Additionally, we observe in
Fig. \ref{bc.fig} that SPT indeed predicts a small dependence of
$b_{\mbox{\scriptsize eff}}c_i$ on $L/D_{\mbox{\scriptsize eff}}$,
in contrast to the $L/D_{\mbox{\scriptsize eff}}$ independent
second virial theory. This suggests that effective aspect ratio of
the rods, which decreases with ionic strength, has a small effect
on the I-N transition concentration. However, the
$L/D_{\mbox{\scriptsize eff}}$ dependence predicted by SPT is
opposite the trend experimentally observed; increasing
$D_{\mbox{\scriptsize eff}}$, by lowering ionic strength,
increases the measured $b_{\mbox{\scriptsize eff}}c_i$ but lowers
the scaled particle theory $b_{\mbox{\scriptsize eff}}c_i$. We
argue that this discrepancy between scaled particle theory and
experimental results at low ionic strength is due to the
approximate treatment of electrostatics in $D_{\mbox{\scriptsize
eff}}$, which is used not only as the theoretical hard rod
diameter in SPT but also scales the experimental coexistence
concentrations from $\rho_i$ to $b_{\mbox{\scriptsize eff}}c_i$.
$D_{\mbox{\scriptsize eff}}$ is determined using from the second
virial coefficient, and therefore is not necessarily accurate
beyond that limit, ie. at low ionic strength. We note that the
rescaled experimental coexistence concentrations,
$b_{\mbox{\scriptsize eff}}c_i$, are extremely sensitive to the
value of $D_{\mbox{\scriptsize eff}}$ used to rescale the measured
coexistence concentrations, $\rho_i$. Differences in
$D_{\mbox{\scriptsize eff}}$ are translated linearly to changes in
the experimental $b_{\mbox{\scriptsize eff}}c_i$ by
$b_{\mbox{\scriptsize eff}}c_i= 25 \rho_i
[\mbox{mg/ml}]L[\mu\mbox{m}] D_{\mbox{\scriptsize eff}}[\mu
\mbox{m}]$. However, the predicted effect of changing
$L/D_{\mbox{\scriptsize eff}}$ on $b_{\mbox{\scriptsize eff}}c_i$,
as shown by the SPT curves in Fig. \ref{bc.fig}, is much smaller
than the measured change in $b_{\mbox{\scriptsize eff}}c_i$ with
ionic strength. Agreement between SPT and our experimental results
improves if the effective diameter at low ionic strength is
smaller than predicted at the second virial limit.

\begin{figure}
\centerline{\epsfig{file=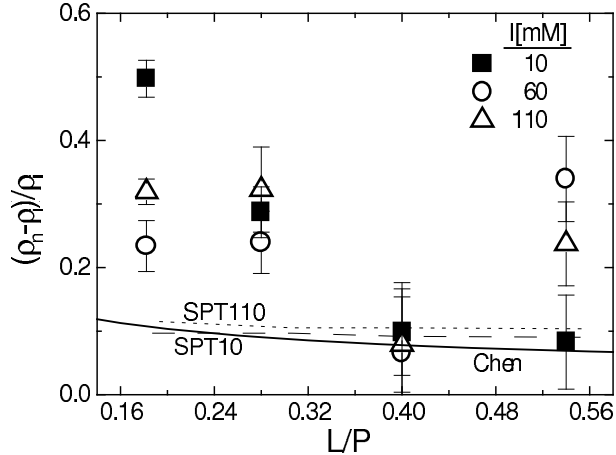,width=8cm}}\caption[Width
of the coexistence region as a function of rod
flexibility]{\label{width.fig} Width, $(\rho_n-\rho_i)/\rho_i$, of
the coexistence region as a function of rod flexibility $L/P$.
Results are plotted for three ionic strengths (10 mM, 60 mM, and
110 mM). Solid line is due to Chen for rods with $P=2.2\mu$m
\cite{Chen93}. Dotted and dashed lines are due to scaled particle
theory for M13 rods (q=7e/nm) with a hard diameter
$D_{\mbox{\scriptsize eff}}$ at 110 mM (SPT110) and 10 mM (SPT10)
ionic strength, respectively. For rigid rods, the Onsager
prediction for the I-N coexistence width is 0.29
~\cite{Onsager49,Herzfeld84}. The width of the coexistence region
should decrease with increasing flexibility.}
\end{figure}

The width of the coexistence region, $(\rho_n -\rho_i)/\rho_i$,
was also measured and is presented in Fig. \ref{width.fig}. At low
ionic strength, the coexistence width qualitatively follows the
decrease expected for increasing flexibility shown by the solid
line due to Chen \cite{Chen93}. For most rod lengths the value for
the coexistence width is larger than predicted by both
Khokhlov-Semenov theory and by scaled particle theory. At short
rod lengths this discrepancy is most likely due to the intrinsic
bidispersity of the suspensions, which acts to widen the
coexistence region \cite{Lekkerkerker84}. A slow increase in the
coexistence concentrations with time (possibly due to bacterial
growth) \cite{Tang93} contributes to the large error bars, making
comparison to predictions difficult. Above 10 mM ionic strength,
where we see strong agreement between measurements of the
coexistence concentrations and theoretical predictions, it is not
apparent that there is any flexibility or ionic strength
dependence in the width measurements. 

\begin{figure}
\centerline{\epsfig{file=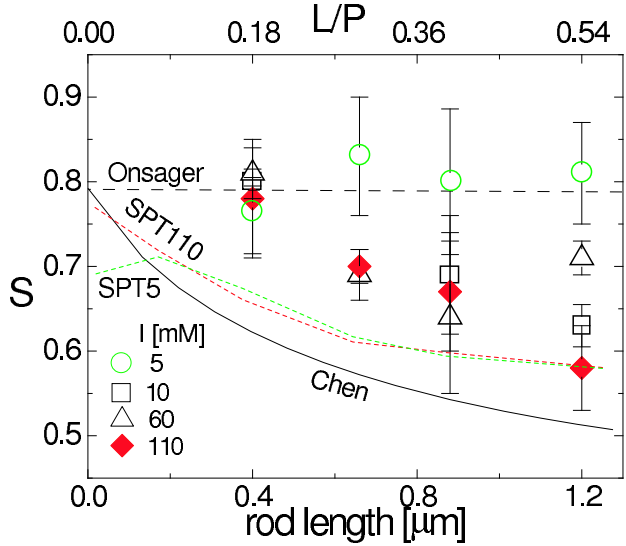,width=8cm}}\caption[Nematic
order parameter ($S$) of the nematic phase in coexistence with the
isotropic phase as a function of particle length.
]{\label{INSmutant.fig}(Color online) Nematic order parameter at
coexistence as a function of rod length for four different ionic
strengths. Solid black line represents the theoretical calculation
by Chen \cite{Chen93} for the order parameter as a function of
flexibility ($L/P$) indicated by the scale on the top of the
graph. The dashed line is the theoretical nematic order parameter
for rigid rods, $S=0.79$\cite{Onsager49,Herzfeld84}. The scaled
particle curves (dotted lines) are calculated as in \cite{Purdy03}
for virus rods at 110 mM (SPT110) and 5 mM (SPT5) ionic strength.
Theoretical curves were calculated for rods with a persistence
length of 2.2$\mu$m. The measured order parameter decreases with
increasing particle length at high ionic strength, but remains
constant at low ionic strength. }
\end{figure}

The nematic order parameter obtained from measurements of the
birefringence of the magnetically unwound and aligned cholesteric
phase in coexistence with the isotropic phase is presented in Fig.
\ref{INSmutant.fig}. We observe that at high ionic strengths, the
nematic order parameter decreases with increasing length
(increasing flexibility) in qualitative agreement with
Khokhlov-Semenov theory calculated by Chen \cite{Chen93}. With
decreasing ionic strength, however, the measured nematic order
parameter increases, approaching Onsager's rigid-rod predictions,
due to increasing the range of electrostatic interactions. This
has also been observed for {\it fd} virus suspensions
\cite{Purdy03}. Furthermore, at very low ionic strength (5 mM
ionic strength) the nematic order parameter becomes independent of
rod length and equal to the predicted rigid rod value of $S=0.8$.
Scaled particle theory, as illustrated in Fig.
\ref{INSmutant.fig}, predicts that the nematic order parameter is
largely independent of ionic strength. This suggests that the
effective aspect ratio of the rods, which decreases with ionic
strength, does not effect the nematic ordering. In addition, SPT
agrees with the experimental measurements at high ionic strength
better than Khokhlov-Semenov theory.

Another possible explanation for an increase in nematic order
parameter with decreasing ionic strength is electrostatic
stiffening. If the interparticle interactions are dominated by
electrostatics, the flexibility of the rods might be screened.
This effective ``electrostatic persistence length" $P_{el}$, which
makes a charged polymer more rigid when in solution, is a dominant
effect in determining the flexibility of charged flexible polymers
with $L/P\gg1$. However, for the semiflexible M13 and {\it fd},
$P_{el}$ is predicted to be less than one percent larger than the
bare persistence length \cite{Odijk78}. Additionally, the results
for the coexistence concentrations presented in Fig. \ref{bc.fig}
indicate that with decreasing ionic strength the measured
coexistence concentrations deviate further from Onsager's
rigid-rod predictions. Thus the measured coexistence
concentrations and nematic order parameters exhibit contradictory
trends, away from Onsager's rigid rod prediction versus towards
Onsager's rigid rod prediction, respectively, with decreasing
ionic strength. Therefore, electrostatic stiffening of the polymer
cannot account for the observed high values of the order parameter
at low ionic strength. Neither scaled particle theory, nor
variation in the electrostatic persistence length satisfactorily
explain the low ionic strength data.

\subsection{Effect of viral surface charge on the isotropic-nematic transition}\label{charge}
\begin{figure}
\centerline{\epsfig{file=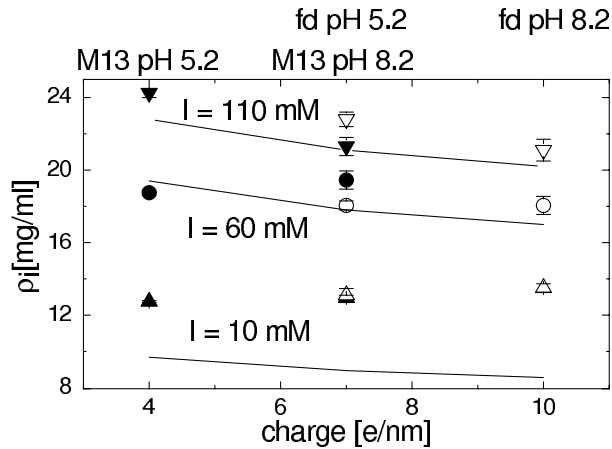,width=8cm}}
\caption[Isotropic coexistence concentration as a function of rod
surface charge. ]{\label{IN.fig} Coexisting isotropic phase
concentrations $\rho_i$ as a function of particle surface charge
for three ionic strengths, 10 mM, 60 mM, and 110 mM. Solid symbols
are wild type M13 and open symbols are {\it fd} suspensions.
Suspension pH is labeled above the graph for M13 and {\it fd}
samples. Solid line is from scaled particle theory for
semiflexible hard rods with a diameter $D_{\mbox{\scriptsize
eff}}$ and $L/P=0.4$. The charge dependence of the phase
transition is well described by theory for ionic strengths of 60
mM and 110 mM.}
\end{figure}

In this section we compare the phase behavior of M13 virus to that
of {\it fd} virus as a function of surface charge and ionic
strength. Recall that these particles have the same length
$L=0.88\mu$m and persistence length $P=2.2\mu$m. In Fig.
\ref{IN.fig} we present measurements of the isotropic coexistence
concentrations as a function of viral surface charge at high and
low ionic strength. The theoretical curve is from scaled particle
theory for charged, semiflexible rods with $L/P=0.4$. We only
present the theoretical results from scaled particle theory in
this section as this theory should more accurately describe the
finite-length rod phase behavior than the second virial theory. In
Fig. \ref{IN.fig} we confirm that the charge dependence of the I-N
coexistence concentrations is accurately described by scaled
particle theory at high ionic strengths. However, the efficacy of
$D_{\mbox{\scriptsize eff}}$ as a means for incorporating all
electrostatic interactions again diminishes at low ionic strength
(I$<$60 mM), as seen previously in Fig. \ref{bc.fig} and in Fig.
~\ref{INSmutant.fig}.

\begin{figure}
\centerline{\epsfig{file=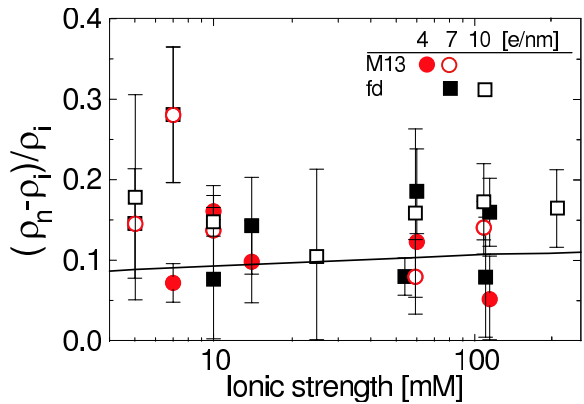,width=8cm}}
\caption[Width of the I-N coexistence region for rods of different
charge. ]{\label{chargewidth.fig} (Color online) Width of the
isotropic-nematic coexistence region for wild type M13 and {\it
fd} rods at three different surface charges as a function of ionic
strength. Both M13 and {\it fd} have a length of $L=0.88\mu$m.
Solid symbols are at pH 5.2 and open symbols are at pH 8.2 for M13
(circles) and {\it fd} (squares) suspensions. Solid line is from
scaled particle theory for hard semiflexible rods with $L/P=0.4$
and is independent of rod surface charge. The Onsager prediction
for the I-N coexistence width in dimensionless units of $bc$ for
hard rigid rods is $(4.19-3.29)/3.29=0.29$
~\cite{Onsager49,Herzfeld84}. The coexistence width does not
clearly show any charge dependence. }
\end{figure}

Fig. \ref{chargewidth.fig} presents the width of the coexistence
region as a function of charge and ionic strength. The width of
the coexistence region is independent of the surface charge of the
rods and agrees (within large error bars) with scaled particle
theory predictions.  Both the measured coexistence concentrations
and coexistence widths show that the effect of surface charge on
the electrostatic interactions which drive the I-N phase
transition are weak, which is consistent with the idea of charge
renormalization incorporated into the calculations of
$D_{\mbox{\scriptsize eff}}$.

\begin{figure}
\centerline{\epsfig{file=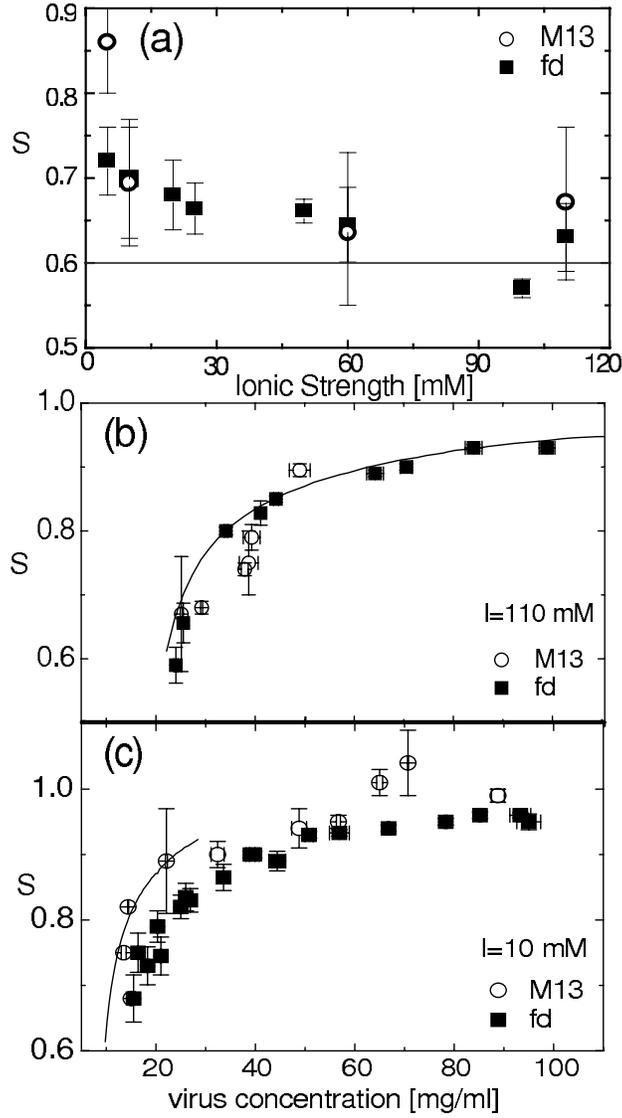,width=8cm}}\caption[Nematic
order parameter as a function of surface charge and ionic
strength. ]{\label{3M13fdS.fig} Order parameter of the nematic
phase (a) coexisting with the isotropic phase as a function of
ionic strength (b)as a function of concentration at 110 mM ionic
strength and (c) as a function of concentration at 10 mM ionic
strength pH 8.2. Values for M13 were obtained by birefringence
measurements and values for {\it fd} were obtained previously by
x-ray diffraction \cite{Purdy03}. Solid lines are scaled particle
theory for semiflexible hard rods of diameter
$D_{\mbox{\scriptsize eff}}$ and $L/P=0.4$. The order parameters
for M13 agree with those measured for {\it fd} independent of
concentration and ionic strength. }
\end{figure}

Nematic order parameters obtained from measurements of the
birefringence of magnetically unwound and aligned cholesteric
samples of M13 at pH 8.2 are compared to previous measurements of
{\it fd} suspension nematic order parameters, measured via x-ray
diffraction techniques, also at pH 8.2 \cite{Purdy03}, in Fig.
\ref{3M13fdS.fig}. Recall that the nematic order parameter of {\it
fd} is known to be proportional to the birefringence of the
suspension by the relationship $S=\Delta n/ \Delta
n_{\mbox{\scriptsize sat}}$ where $\Delta
n_{\mbox{\scriptsize{sat}}}=3.8 \times 10^{-5}$ml/mg
\cite{Purdy03}. The order parameter of M13 was measured at I-N
coexistence as a function of ionic strength, and deep within the
nematic phase for high (110 mM) and low (10 mM) ionic strength.
Theoretical predictions from scaled particle theory for the
nematic order parameter of hard semiflexible rods with $L/P=0.4$
are also shown in Fig. \ref{3M13fdS.fig}. The order parameters of
M13 and {\it fd} were found to be equal as a function of ionic
strength and concentration, indicating that the surface charge
difference of 30\% between the two particles does not effect
nematic ordering. The insensitivity of the nematic order parameter
to surface charge is consistent with the surface charge
renormalization incorporated into $D_{\mbox{\scriptsize eff}}$
calculations (Fig. \ref{Deff.fig}) \cite{Tang95}. The strong
agreement of M13 and {\it fd} order parameters also indicates that
these two different virus particles have the same birefringence
per particle, $\Delta n_{\mbox{\scriptsize{sat}}}=3.8 \times
10^{-5}$ml/mg \cite{Purdy03}. Additionally, we again observe that
the scaled particle theory fits the measured order parameter best
for high ionic strength data.

\section{Conclusion}
At high ionic strengths, where the range of electrostatic
interactions are small and $L/D_{\mbox{\scriptsize eff}}$ is
large, the isotropic-nematic transition of the experimental system
of charged semiflexible bacteriophages is well described by
Khokhlov-Semenov theory for semiflexible charged rods. Increasing
flexibility increases the coexistence concentrations
$b_{\mbox{\scriptsize eff}}c_i$ (Fig. \ref{bc.fig}) and lowers the
nematic order parameter (Fig. \ref{INSmutant.fig}). In the region
of high ionic strength, $D_{\mbox{\scriptsize{eff}}}$ accurately
describes both the charge dependence and ionic strength dependence
of the isotropic-nematic phase transition (Fig. \ref{IN.fig}). At
low ionic strength, however, we find that the I-N coexistence
concentrations and the nematic order parameter do not agree with
theoretical predictions from either Onsager's second virial
theory, or scaled particle theory. At low ionic strength, the
flexibility dependence of the nematic order parameter is much
weaker than expected (Fig. \ref{INSmutant.fig}), but the
flexibility dependence the coexistence concentrations is much
stronger than expected (Fig. \ref{bc.fig}). Because of these
contradictory results we suggest that the disagreement between
theoretical predictions and experimental data at low ionic
strength is due to the approximate incorporation of the
electrostatic interactions into the theoretical free energy via
$D_{\mbox{\scriptsize eff}}$.

\begin{acknowledgments}
We would like to thank Zvonimir Dogic for the program which
calculates the scaled particle theory phase diagram. We also
acknowledge support from the National Science Foundation (DMR-CMP
0088008).
\end{acknowledgments}

\end{document}